\documentclass[english,aps,prl,twocolumn,showpacs,superscriptaddress,groupedaddress,footinbib]{revtex4}  % for review and submission
\usepackage{graphicx}  % needed for figures
\usepackage{dcolumn}   % needed for some tables
\usepackage{bm}        % for math
\usepackage{amssymb}   % for math
\usepackage{babel}
\usepackage{verbatim}
\usepackage{amsmath}
\usepackage{setspace}
 \usepackage{epstopdf}

\graphicspath{{figures/}}

\usepackage{siunitx}

\begin{document}

 %The following information is for internal review, please remove them for submission
\widetext

\title{Experimental realization of time-dependent phase-modulated continuous dynamical decoupling}

\author{D. Farfurnik}
\affiliation{Racah Institute of Physics, The Hebrew University of Jerusalem, Jerusalem 9190401, Israel}
\affiliation{The Center for Nanoscience and Nanotechnology, The Hebrew University of Jerusalem, Jerusalem 9190401, Israel}

\author{N. Aharon}
\affiliation{Racah Institute of Physics, The Hebrew University of Jerusalem, Jerusalem 9190401, Israel}

\author{I. Cohen}
\affiliation{Racah Institute of Physics, The Hebrew University of Jerusalem, Jerusalem 9190401, Israel}

\author{Y. Hovav}
\affiliation{Dept. of Applied Physics, Rachel and Selim School of Engineering, Hebrew University, Jerusalem 9190401, Israel}

\author{A. Retzker}
\affiliation{Racah Institute of Physics, The Hebrew University of Jerusalem, Jerusalem 9190401, Israel}

\author{N. Bar-Gill}
\affiliation{Racah Institute of Physics, The Hebrew University of Jerusalem, Jerusalem 9190401, Israel}
\affiliation{The Center for Nanoscience and Nanotechnology, The Hebrew University of Jerusalem, Jerusalem 9190401, Israel}
\affiliation{Dept. of Applied Physics, Rachel and Selim School of Engineering, Hebrew University, Jerusalem 9190401, Israel}

%\input author_list.tex       % D0 authors (remove the first 3 lines
                             % of this file prior to submission, they
                             % contain a time stamp for the authorlist)
                             % (includes institutions and visitors)
\date{\today}
\begin{abstract}
The coherence times achieved with continuous dynamical decoupling techniques are often limited by fluctuations in the driving amplitude. In this work, we use time-dependent phase-modulated continuous driving to increase the robustness against such fluctuations in a dense ensemble of nitrogen-vacancy centers in diamond. Considering realistic experimental errors in the system, we identify the optimal modulation strength, and demonstrate an improvement of an order of magnitude in the spin-preservation of arbitrary states over conventional single continuous driving. The phase-modulated driving exhibits comparable results to previously examined amplitude-modulated techniques, and is expected to outperform them in experimental systems having higher phase accuracy. The proposed technique could open new avenues for quantum information processing and many body physics, in systems dominated by high-frequency spin-bath noise, for which pulsed dynamical decoupling is less effective. 

\end{abstract}

\maketitle

\begin{figure}[!t]	
	\includegraphics[width=1.05\columnwidth]{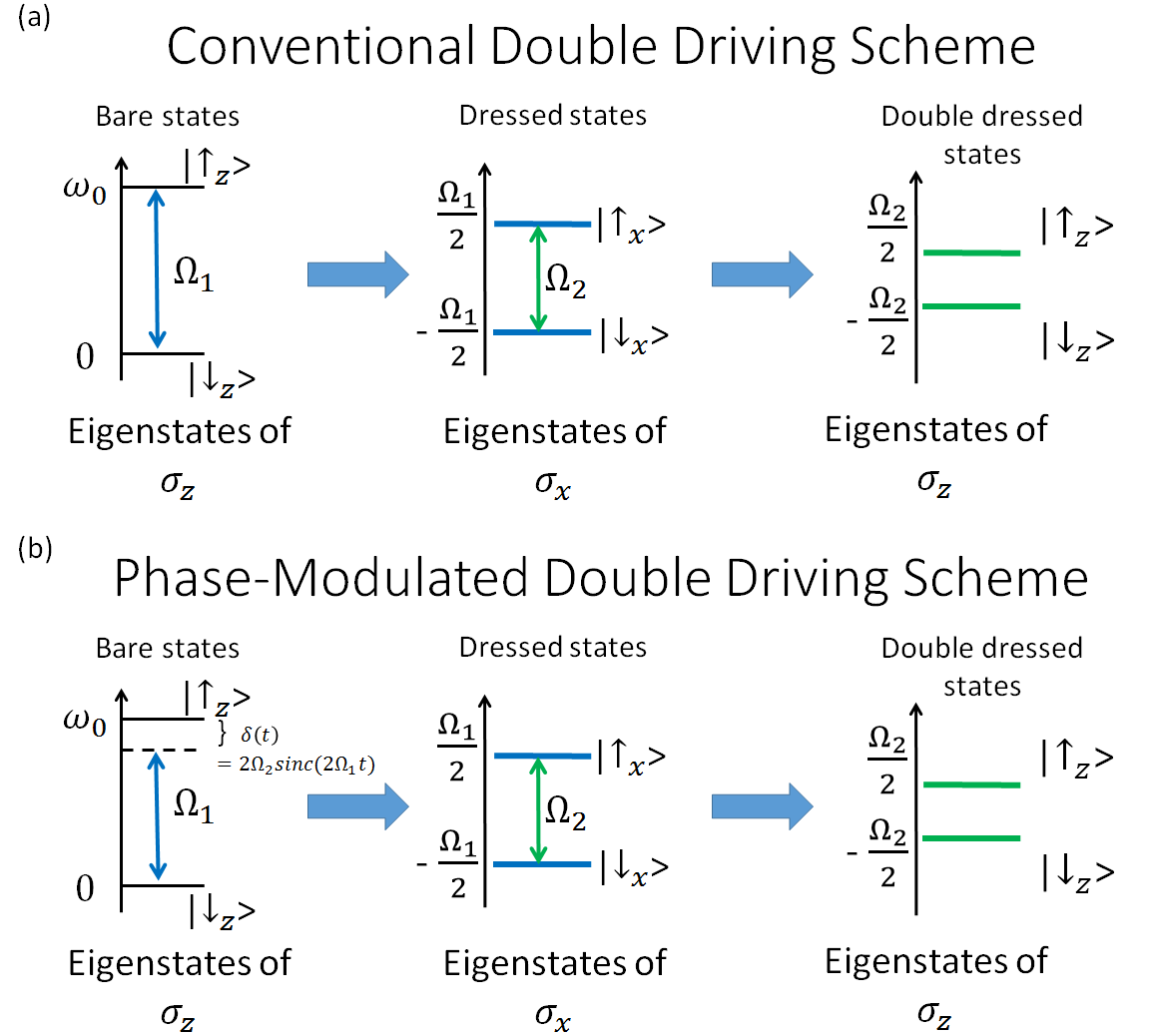}
	\caption{(Color online) (a) Conventional double driving scheme, utilizing two microwave sources: first driving with amplitude $\Omega_1$ along the x axis and second driving with amplitude $\Omega_2$ along the z axis. For $\Omega_2<\Omega_1$, the effective Hamiltonian (\ref{equation:HI2}) is reproduced, significantly reducing amplitude fluctuations of the first driving. (b) Phase-modulated double driving scheme, utilizing a single microwave source, and time-dependent phase modulation $\delta(t)=2\Omega_2 sinc(2\Omega_1 t)$, reproducing the same effective Hamiltonian (\ref{equation:HI2}) (appendix A).} 
	\label{fig:DoubleDrivingScheme}
\end{figure}

%%%%%%%%%%%%%%%%%%%%%%%%%%%%%%%%%%%%%%%%%%%%%%%%%%%%%%
%% INTRODUCTION
%%%%%%%%%%%%%%%%%%%%%%%%%%%%%%%%%%%%%%%%%%%%%%%%%%%%%%
\paragraph{}
One of the main challenges in quantum information processing, quantum computing and quantum sensing is the preservation of arbitrary spin states. For example, the sensitivity of nitrogen-vacancy (NV) ensemble-based AC magneteometry scales as a square-root of the coherence time  \cite{Taylor2008, Maze2008, Balasubramanian2008, Grinolds2011, Pham2011, Pham2012, Acosta2009, Acosta2010}. Moreover, long ensemble spin coherence times could open new avenues for studying many body dynamics of interacting spins \cite{Cappellaro2009,Bennett2013,Weimer2013}. The commonly used technique for improving coherence times and preserving arbitrary states is dynamical decoupling (DD) sequences \cite{Ryan2010,Naydenov2011,Shim2012,Hirose2012,Cai2012,Farfurnik2015,Teissier2017}. Although pulsed DD is very efficient for a variety of physical systems, continuous driving-based decoupling (i.e. spin-lock) has an advantage when the power spectrum of the noise bath contains a significant contribution from high-frequency terms, such that relevant correlation times are shorter than the duty cycle achievable by pulsed techniques \cite{Hirose2012,Cai2012}. However, in these continuous schemes, amplitude fluctuations of the driving source itself limit the achieved coherence times, raising the need for a fault tolerant driving \cite{Solomon1959,Timoney2011,Cai2012,Aharon2013,Cohen2016,Aharon2016,Teissier2017}. 
\paragraph{}
One common approach for overcoming fluctuations in the driving amplitude is to flip the phase of the continuous driving every time increment $\tau$ (i.e., to apply a ``rotary echo", \cite{Aiello2013}). However, in the basis of the dressed states, these techniques are equivalent to pulsed DD, having the same disadvantages, such as additional imperfections due to the application of non-ideal pulses, and the disability of mitigating amplitude fluctuations that are faster than the flipping rate $1/\tau$. Another approach for overcoming these fluctuations is to apply an additional continuous driving in the perpendicular axis [Fig. \ref{fig:DoubleDrivingScheme}(a)]. In order to avoid the use of an extra microwave (MW) source, the same effective Hamiltonian can be generated by a time-dependent modulation of the amplitude or phase of the original driving. Recently, such time-dependent amplitude modulation was experimentally demonstrated in a system of single isolated NV centers, achieving an order of magnitude improvement in the coherence time over a single continuous driving \cite{Cai2012}. 
\paragraph{}
In this work, we demonstrate the effectiveness of the same approach for a dense ensemble of NVs, and implement a technically advantageous time-dependent phase modulation approach [Fig. \ref{fig:DoubleDrivingScheme}(b)] \cite{Cohen2016}, producing a similar order of magnitude improvement in the coherence times. By additionally considering the effect of amplitude/phase modulation on the preservation of the spin component along the driving axis, and by taking into account the reduction of the fluorescence signal contrast due to the modulation, we identify the optimal modulation strength for the preservation of arbitrary spin states of the NV ensemble.
\paragraph{}
We describe the non-interacting NV ensemble in our system as an effective two level system under continuous driving. The Hamiltonian of the system is given by
\begin{equation}
H=\frac{\omega_0}{2} \sigma_z + f(t)  \sigma_z +\Omega_1 \cos(\omega_0 t)\sigma_x,
\end{equation} where $\omega_0$ is the energy of the $|0\rangle\leftrightarrow|1\rangle$ transition, $f(t)$ represents the effective time-dependence of the spin-bath noise, and $\Omega_1$ is the continuous driving amplitude (Rabi frequency). Moving to the rotating frame with respect to $H_0=\frac{\omega_0}{2} \sigma_z$, under the rotating wave approximation yields
 \begin{equation}
 H_I=\frac{\Omega_1}{2} \sigma_x + f(t)  \sigma_z.
 \end{equation} For Rabi frequencies much larger than the significant frequencies in the power spectrum of the noise, the second term in the Hamiltonian is greatly diminished, leading to the decoupling of the spin-bath, thereby increasing the coherence time of a spin state initialized along the perpendicular ($y$) axis. Experimentally, however, fluctuations in the applied driving amplitude $\delta \Omega_1(t)$ will eventually limit the achieved coherence times \cite{Cai2012,Cohen2016,Aharon2016}.
 In order to overcome the effect of such fluctuations, an additional driving with a frequency $\Omega_1$ and an amplitude $\Omega_2$ can be applied along the $z$ axis [Fig. \ref{fig:DoubleDrivingScheme}(a)]. The resulting effective Hamiltonian in the second interaction picture, now with respect to $H_1=\frac{\Omega_1}{2} \sigma_x$, yields
   \begin{equation}
   H_{I2}=\frac{\Omega_2}{2} \sigma_z + \frac{\delta \Omega_1(t)}{2} \sigma_x.
   \label{equation:HI2}
   \end{equation} For a small amplitude of the second driving, $\frac{\Omega_2}{\Omega_1}<1$, and assuming that the amplitude fluctuations from the MW source scale accordingly, $\delta \Omega_2=\frac{\Omega_2}{\Omega_1}\delta\Omega_1$, the remaining fluctuations $\delta \Omega_2 < \delta \Omega_1$ will result in further enhancement of the coherence times. In fact, such an implementation does not require an additional MW source. By applying a single drive, either with time-dependent amplitude modulation of the form [Fig. \ref{fig:DoubleDrivingScheme}(a), same technique used in \cite{Cai2012}]
           \begin{equation}
           \begin{split}
            & \Omega_1 [\cos(\omega_0 t)+\alpha \sin(\Omega_1 t) \cos(\omega_0 t +\frac{\pi}{2})] \sigma_x =\\ 
           &= \Omega_1 \cos(\omega_0 t)[ \sigma_x+\alpha \sin(\Omega_1 t)\sigma_y],
           \label{equation:ampdetun}
           \end{split}
      \end{equation} or with time-dependent phase modulation of the form [Fig. \ref{fig:DoubleDrivingScheme}(b)],
           \begin{equation}
           \begin{split}
           &\Omega_1 \cos[\omega_0 t+\alpha \sin(\Omega_1 t)] \sigma_x=\\
           &=\Omega_1 \cos(\omega_0 t) \{\cos[\alpha \sin(\Omega_1 t)]\sigma_x+\sin[\alpha \sin(\Omega_1 t)]\sigma_y\},
           \label{equation:phasedetun}
           \end{split}
      \end{equation} where $\alpha\equiv 2\frac{\Omega_2}{\Omega_1}$ is the ``modulation strength", the form of the Hamiltonian (\ref{equation:HI2}) is reproduced  (appendix A). Experimentally, such drivings can be produced by an arbitrary waveform generator (AWG). If the timing resolution of the AWG is much larger than $\omega_0$, the waveforms  (\ref{equation:ampdetun}),(\ref{equation:phasedetun}) can be programmed explicitly. In our case, however, the timing resolution was limited to 1 ns, and the modulation of the amplitude/phase was created through in-phase-and-quadrature (I/Q) mixing: two (``I" and ``Q") waveforms of the I/Q mixer were introduced from two separate channels of the AWG as waveforms, $[1,\alpha \sin(\Omega_1 t)]$ for the amplitude modulation scheme and $\{\cos[\alpha \sin(\Omega_1 t)],\sin[\alpha \sin(\Omega_1 t)]\}$ for the phase modulation scheme.
 
%%%%%%%%%%%%%%%%%%%%%%%%%%%%%%%%%%%%%%%%%%%%%%%%%%%%%%
%% Methods
%%%%%%%%%%%%%%%%%%%%%%%%%%%%%%%%%%%%%%%%%%%%%%%%%%%%%%

\paragraph{}
We performed measurements on an isotopically pure ($99.99\%$ $^{12}C$) diamond sample with nitrogen concentration of $\sim 2 \times 10^{17}$ $\SI{}{\centi\meter}^{-3}$  and NV concentration of $\sim 4 \times 10^{14}$ $\SI{}{\centi\meter}^{-3}$, grown via chemical vapor deposition (Element Six). A $532$ nm laser was used to induce fluorescence from $\sim 10^4$ NV centers within a $\sim 20$ $\SI{}{\micro\meter} ^3$ measurement volume, and the resulting fluorescence signal was measured using a single photon counting module. After Zeeman-splitting the ground state energy levels using a $\sim 40$ Gauss permanent magnet, continuous driving at a Rabi frequency $\Omega_1\approx 9$ $\SI{}{\mega\hertz}$ was introduced resonantly with the $|0\rangle \leftrightarrow |1\rangle$ transition (SRS SG384 signal generator). The time-dependent phase/amplitude of the driving was generated using I/Q modulation, produced by two output channels of an AWG with 1 GHz sampling rate (Agilent 33621A). Using a fast oscilloscope, we experimentally estimated the relative amplitude fluctuations as $\Delta A/A\approx0.75\%$ and phase fluctuations as $\Delta \phi\approx7$ mrad. 

\paragraph{}
We first measured the decoherence of the ensemble spin state in the presence of a single driving field without time-dependent modulations (equivalent to a typical ``Rabi" driving experiment). A resonant $\frac{\pi}{2}$-pulse, initializing the state along the y axis, was followed by driving along the x axis. The state was then rotated onto the z-axis of the Bloch sphere using another $\frac{\pi}{2}$-pulse, and the fidelity of the state was determined from the fluorescence signal contrast (appendix D). The decoherence curve was extracted by scanning the total driving time. We then repeated these measurements while introducing amplitude and phase dependent modulations, according to equations (\ref{equation:ampdetun}),(\ref{equation:phasedetun}).

\begin{figure}[!t]	
	\includegraphics[width=1.05\columnwidth]{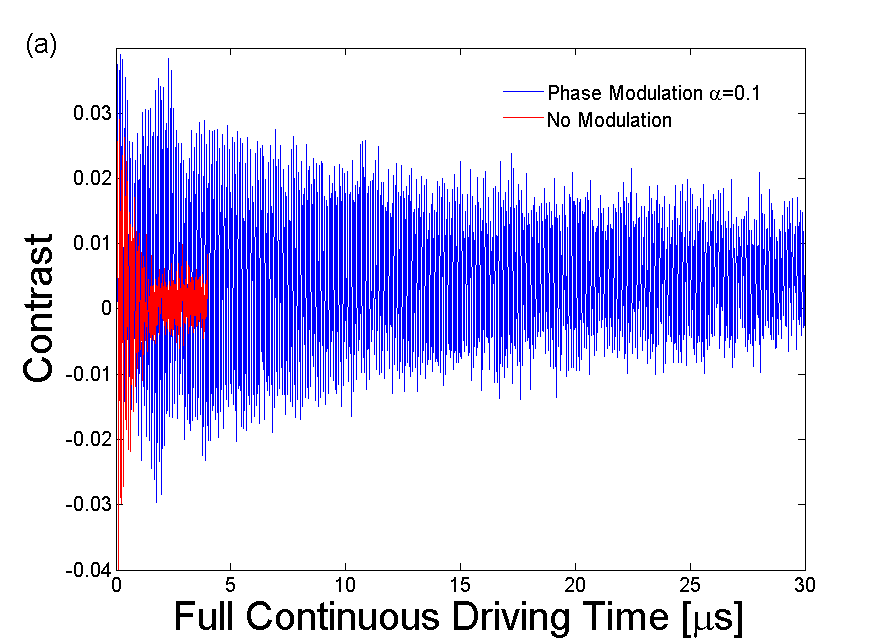} %fig3a.eps
	\includegraphics[width=1.05\columnwidth]{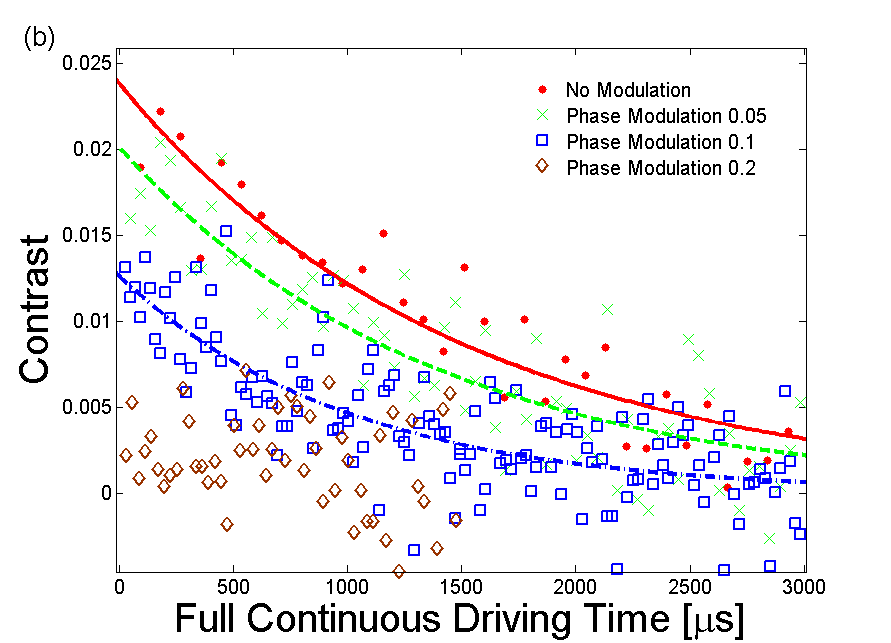} %fig3b.eps
	\caption{(Color online) Fidelity of the ensemble spin state (appendix D) after performing continuous driving along the $x$ axis.  (a) Conventional continuous driving with no modulation, and double driving with phase modulation strength $\alpha=0.1$, for a state initialized along the $y$ axis ($T_2$ decay with ``Rabi" oscillations). (b) Double driving with different phase modulation strengths for a state initialized along the $x$ axis ($T_{1\rho}$ decay).} 
	\label{fig:driving}
\end{figure}

\begin{figure}[!t]	
	\includegraphics[width=1.05\columnwidth]{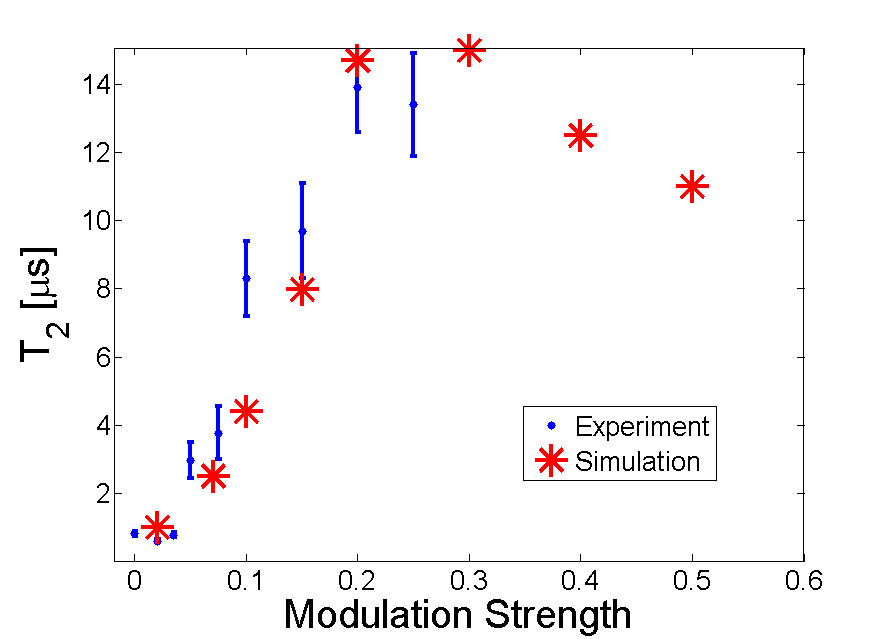} %fig3a.eps
	\caption{(Color online) Decoherence time of the ensemble spin state after performing phase-modulated continuous driving, as a function of the phase modulation strength. The state was initialized along the y axis ($T_2$ decay).} 
	\label{fig:T2scaling}
\end{figure}

%%%%%%%%%%%%%%%%%%%%%%%%%%%%%%%%%%%%%%%%%%%%%%%%%%%%%%
%% Results
%%%%%%%%%%%%%%%%%%%%%%%%%%%%%%%%%%%%%%%%%%%%%%%%%%%%%%
 In Fig. \ref{fig:driving}(a) we demonstrate an order of magnitude improvement in the coherence time, from $T_2=0.81$ $\SI{}{\micro\second}$ using conventional single continuous driving with no modulation (``Rabi" experiment) up to $T_2=8.3$ $\SI{}{\micro\second}$ using phase modulation with a strength of $\alpha=0.1$ [eq. (\ref{equation:phasedetun})]. 
Note that the spin state does not decay to zero in the timescale decipated in the plot, due to the existence of an off-resonant hyperfine term, which decays on the longitudinal relaxation timescale $T_1$. By performing a similar analysis, we extracted the coherence time for different modulation strengths (Fig. \ref{fig:T2scaling}). There is a good agreement between the experimental results and a simulation of decoherence considering Orenstein-Uhlenbeck spin-bath model and continuous driving with time-dependent phase modulation (appendix B).  On the one hand, larger values of $\alpha$ are required to increase the coherence times in the presence of large inhomogeneous broadenings, which are significant for spin ensembles. On the other hand, fluctuations in the second driving $d \Omega_2$ still remain, and although they are much weaker than the fluctuations in the first driving, if the modulation strength $\alpha$ is too large, these second-order fluctuations will eventually limit the coherence times. Considering these effects, given the inhomogeneities, spin-bath noise and amplitude fluctuations in our system, a maximal coherence time of $T_2\approx 15$ $\SI{}{\micro\second}$ is predicted by simulations, with a modulation strength of $\alpha\approx 0.3$.
\paragraph{}
Similar results were obtained in the case of amplitude modulation [eq. (\ref{equation:ampdetun})] with equivalent strengths (appendix C), in agreement with previous observations with the exact same technique on single isolated NV centers \cite{Cai2012}. The choice between phase and amplitude modulation is uniquely motivated by experimental considerations. Theoretically, both approaches yield the same Hamiltonian expressed in eq. (\ref{equation:HI2}). Experimentally, however, the limiting factor for extending the coherence time is the remaining fluctuations in the modulated parameter (amplitude / phase). In our case, since both amplitude and phase originate from the same experimental source (two identical channels of the same AWG), the results are very similar. However, we expect that by using an AWG with faster sampling rate, allowing the full modulated waveform to be directly programmed, much higher phase accuracy could be achieved. In such a case, the phase modulation approach is expected to outperform amplitude modulation for equivalent modulation strengths. The order of magnitude improvement in the coherence times is also consistent with the results for single NVs \cite{Cai2012}, with the only difference being that here, much larger modulation strengths are needed due to the inhomogeneities over the ensemble's measurement volume.

\begin{figure}[!t]	
	\includegraphics[width=1.05\columnwidth]{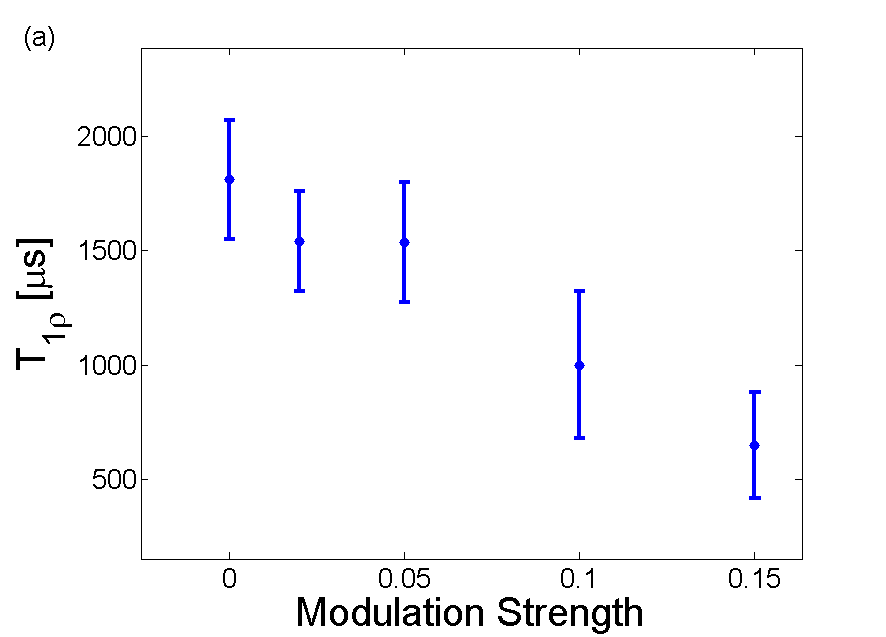} 
	\includegraphics[width=1.05\columnwidth]{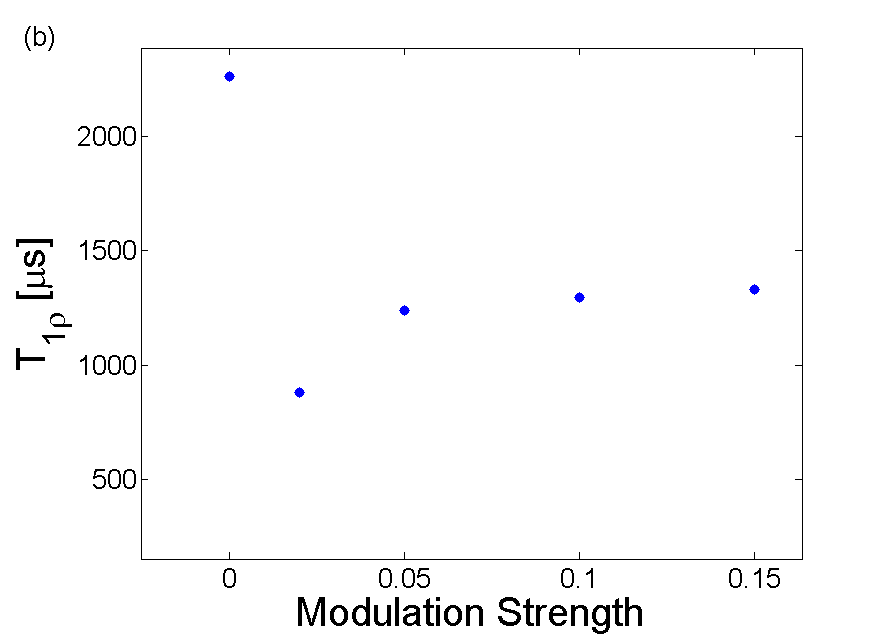} 
		\includegraphics[width=1.05\columnwidth]{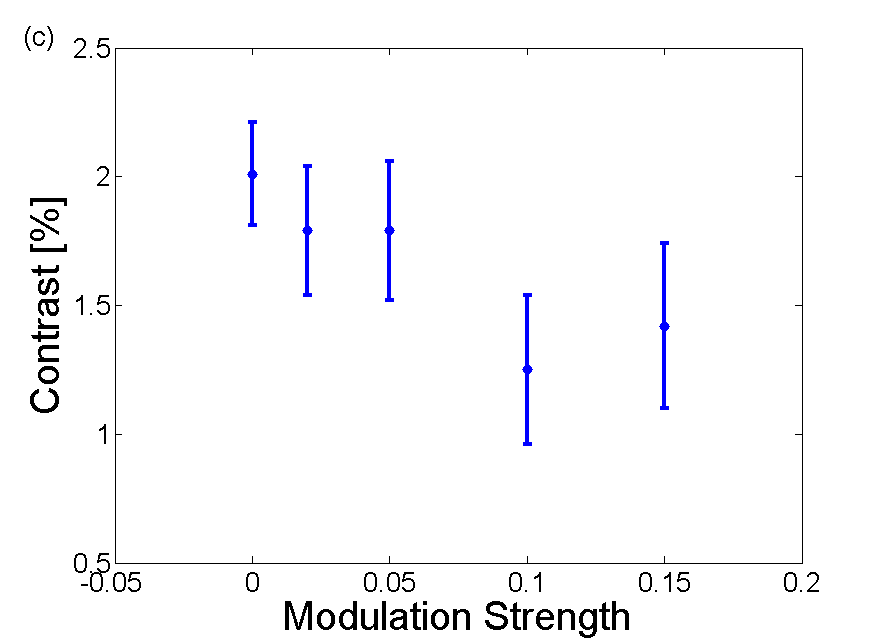} 
	\caption{(Color online) (a) Experimental results and (b) simulations of the relaxation time of the spin ensemble state after performing phase-modulated continuous driving as a function of the phase modulation strength. The state was initialized along the driving axis. (c) Initial fluorescence signal contrast measured in the same experiment.} 
	\label{fig:T1scaling}
\end{figure}

\paragraph{}
Since we are interested in preserving arbitrary states (and not just a particular state perpendicular to the driving axis), we also performed the same experiment with the spin state initialized along the driving axis (x) [Fig. \ref{fig:driving}(b)]. Without introducing modulation, this experiment represents conventional spin-locking, resulting in a measured relaxation time of $T_{1\rho}\approx1800$ $\SI{}{\micro\second}$, compared to a longitudinal relaxation of $T_{1}\approx5900$ $\SI{}{\micro\second}$. In this case, since phase/amplitude fluctuations are not completely canceled, and due to the discrete sampling of the AWG, the application of phase/amplitude modulation reduces the efficiency of the spin locking, resulting in a simultaneous reduction of the relaxation time $T_{1\rho}$, as well as of the absolute measured fluorescence contrast [Fig. \ref{fig:driving}(b)]. This degradation is plotted in Fig. \ref{fig:T1scaling} as a function of modulation strength: at a modulation strength of $\alpha=0.1$ both relaxation time and signal contrast drop by a factor of $\lesssim 2$, and for modulations larger than $\alpha=0.2$, the contrast drops below the noise floor level [Fig. \ref{fig:driving}(b)], making such modulations irrelevant for the preservation of arbitrary spin states. Note that the experimental results in Fig. \ref{fig:T1scaling} do not completely agree with the simulation results. In particular, while simulations predict saturation in the relaxation time, experimenal results demonstrate degradation in this parameter as a function of the modulation strength. We believe that this disagreement is caused by phase fluctuations in the driving, as well as inhomogeneities in such fluctuations over the measurement volume, which are not properly taken into account in the theoretical model, and will be studied in a future work.  
\paragraph{}
For general applications in quantum information processing, quantum computing and quantum sensing, one needs to preserve an arbitrary state of the spin ensemble, while maintaining a reasonably high signal contrast, which will be applicable for measurements. As a result, for each physical system, the modulation strength $\alpha$ has to be optimized to achieve the highest possible coherence time  with minimal loss of signal contrast for states initialized along the driving axis. In our system, a value of $\alpha=0.1$ offers a good compromise, increasing the off-axis coherence time by an order of magnitude  up to $T_2\approx 8$ $\SI{}{\micro\second}$, while shortening the on-axis relaxation time ($T_{1\rho}\approx 1000$ $\SI{}{\micro\second}$) and contrast ($1.26\%$) by less than a factor of 2 from a conventional spin-lock.  

%%%%%%%%%%%%%%%%%%%%%%%%%%%%%%%%%%%%%%%%%%%%%%%%%%%%%%
%% Summary & Conclusions
%%%%%%%%%%%%%%%%%%%%%%%%%%%%%%%%%%%%%%%%%%%%%%%%%%%%%%
\paragraph{}
To summarize, we have shown that time-dependent phase modulation can improve the coherence time of a dense ensemble of NVs by more than an order of magnitude over conventional continuous DD. In our system, using a modulation strength of $\alpha=0.1$, the coherence of any arbitrary spin state of the ensemble is preserved up to $T_2\approx 8$ $\SI{}{\micro\second}$, with a degradation of less than a factor of two in the contrast of a state initialized along the driving axis. These resulting coherence times are comparable to those achieved by previously studied amplitude modulation, and can be further improved for an experimental implementation with AWG [Fig. \ref{fig:DoubleDrivingScheme}(a)] having higher timing resolutions. The obtained results could potentially be useful for overcoming decoherence due to spin-bath environments having significant high-frequency terms, where pulsed DD schemes are no longer effective. In such conditions, the coherence times achieved in this work could significantly contribute to sensing applications, as well as the study of many body spin physics.

\section*{Acknowledgements}
This work has been supported in part by the Minerva ARCHES award, the CIFAR-Azrieli global scholars program, the Israel Science Foundation (grant No. 750/14), the Ministry of Science and Technology, Israel, and the CAMBR fellowship for Nanoscience and Nanotechnology.  A. R. acknowledges the support of the Israel Science Foundation (grant no. 1500/13), the European commission, EU Project DIADEMS and Hyperdiamond and the Niedersachsen-Israeli Research Cooperation Program.

\appendix
\section{Appendix A: Derivation of the double driving scheme}
Let us consider a two level system with an energy gap $\omega_0$ and random spin-bath noise $f(t)$ along $\sigma_z$, under a continuous driving with amplitude $\Omega_1$ and time-dependent phase modulation of the form $\phi(t)=2\frac{\Omega_2}{\Omega_1} \sin(\Omega_1 t)$. If the fluctuation in the driving amplitude is $\delta \Omega_1(t)$, the Hamiltonian of the system takes the form
 \begin{equation*}
 H=\frac{\omega_0}{2} \sigma_z + f(t)  \sigma_z +[\Omega_1+\delta \Omega_1(t)] \cos[\omega_0 t+\phi(t)]\sigma_x.
 \end{equation*}
 By moving to the rotating frame, with respect to $H_0=\frac{\omega_0}{2} \sigma_z$, the effective Hamiltonian yields $H_I=U^{\dagger}HU-H_0$, with the evolution operator $U=e^{-iH_0t}$. Since $U$ commutes with $\sigma_z$, only the $\sigma_x$ term is affected by $U^{\dagger}HU$:
  \begin{equation*}
   \begin{split}
  &  \cos[\omega_0 t+\phi(t)] e^{i\frac{\omega_0}{2}t}\sigma_xe^{-i\frac{\omega_0}{2}t} =\\ &
 \cos[\omega_0t+\phi(t)] [\cos(\omega_0t)\sigma_x-\sin(\omega_0t)\sigma_y]= \\&
  \frac{1}{2}\{\cos[\phi(t)]+\cos[(2\omega_0t+\phi(t)]\}\sigma_x-\\&
  \frac{1}{2}\{[\sin[\phi(t)]+\sin(2\omega_0t+\phi(t))]\}\sigma_y.
   \end{split}
  \end{equation*}
  In the rotating wave approximation, the rotating terms can be neglected. In the limit $\Omega_2\ll\Omega_1$ we get $\cos[\phi(t)]\rightarrow 1$, $\sin[\phi(t)]\rightarrow \phi(t)=2\frac{\Omega_2}{\Omega_1}\sin(\Omega_1 t)$. Under these assumptions, the Hamiltonian in the first interaction picture yields
  \begin{equation*}
 H_I=\frac{\Omega_1+\delta \Omega_1(t)}{2} \sigma_x - [\Omega_2+\delta\Omega_2(t)] \sin(\Omega_1 t)\sigma_y + f(t)  \sigma_z.
  \end{equation*}
  If the Rabi frequency $\Omega_1$ is much larger than the dominant frequencies of the spin-bath, the last term is greatly diminished.  By moving to the rotating frame, with respect to $H_0=\frac{\Omega_1}{2} \sigma_x$, the remaining part of the first term is $\frac{\delta \Omega_1(t)}{2} \sigma_x$. The second term yields
    \begin{equation*}
    \begin{split}
    &  -\sin[\Omega_1 t] e^{i\frac{\Omega_1}{2}t}\sigma_ye^{-i\frac{\Omega_1}{2}t} =\\ &
   \sin[\Omega_1 t] [\sin(\Omega_1t)\sigma_z-\cos(\Omega_1t)\sigma_y]= \\&
    \frac{1}{2}\{[1-\cos(2\Omega_1t)]\sigma_z-\sin(2\Omega_1t)\sigma_y].
    \end{split}
    \end{equation*} 
    In the rotating wave approximation, the rotating terms can be neglected, and the Hamiltonian in the second interaction picture yields
      \begin{equation*}
      H_{I2}=\frac{\Omega_2+\delta \Omega_2(t)}{2} \sigma_z + \frac{\delta \Omega_1(t)}{2}\sigma_x,
      \end{equation*}
  which in the case of small $\delta \Omega_2$ yields to Eq. \ref{equation:HI2}.
\section{Appendix B: Simulation details}
\paragraph{}
In order to simulate the decoherence of the spin state under continuous driving, the initial state taken into account was along the axis perpendicular to the driving (the ``y" axis). Due to the hyperfine coupling with nitrogen nuclear spins, $1/3$ of the initial state was detuned from the driving frequency by $\sim \pm 2.2 $ MHz, where the width of each peak was set to $1$ MHz (Normally distributed). The decoherence curves were simulated under the evolution of the Hamiltonian in the first interaction picture (appendix A), where both the spin bath-noise $f(t)$, and amplitude fluctuations $\delta \Omega_1$, $\delta \Omega_2$ were modeled as  Ornstein-Uhlenbeck (OU) processes \cite{Wang1945,Hanson2008}. The OU processes had a zero expectation value and a correlation function of $\langle f(t)f(t')\rangle=\frac{c\tau}{2}e^{-\frac{|t-t'|}{\tau}}$, where $\tau$ is the correlation time of the noise and $c$ is the diffusion coefficient. For both spin-bath and amplitude fluctuations, the OU process was realized by an exact algorithm \cite{Gillespie1996}, given by the time propagation $f(t+\Delta t)=f(t)e^{-\frac{\Delta t}{\tau}} +n\sqrt{\frac{c\tau}{2}(1-e^{-\frac{2\Delta t}{\tau}})}$ for a time step of $\Delta t$ and a unit Gaussian random number $n$. For the spin-bath noise $f(t)$, we used a correlation time of $\tau_B=10$ $\mu$s and a diffusion coefficient of $c_B\approx6.6667\times10^{-5}$ MHz$^{3}$, estimated from the nitrogen (P1) defect concentration in the sample (in accordance with a pure dephasing time of $T_2^{*}=200$ ns and a $T_2=300$ $\mu$s by a Hahn Echo pulse). For the amplitude fluctuations, we chose $\tau_{\Omega}=500$ $\mu$s and set $c_{\Omega}= \frac{2(\delta_{\Omega}\Omega)^2}{\tau_{\Omega}}$, where $\delta_{\Omega}=0.75\%$ is the relative amplitude error, which was determined experimentally. Spin-dynamic simulations under this evolution led to theoretical decoherence curves similar to Fig. \ref{fig:driving}(a). 
\paragraph{}
The simulations were  performed for different modulation strengnths $\alpha$, and typical decay times $T_2$ were extracted for each $\alpha$ by a simple exponential fitting. The red dots in Fig. \ref{fig:T2scaling} represent simulation results for these decoherence times at various modulation strengths, in agreement with the experimental results.

\section{Appendix C: Phase versus amplitude modulation results}
\begin{figure}[!t]	
	\includegraphics[width=1.05\columnwidth]{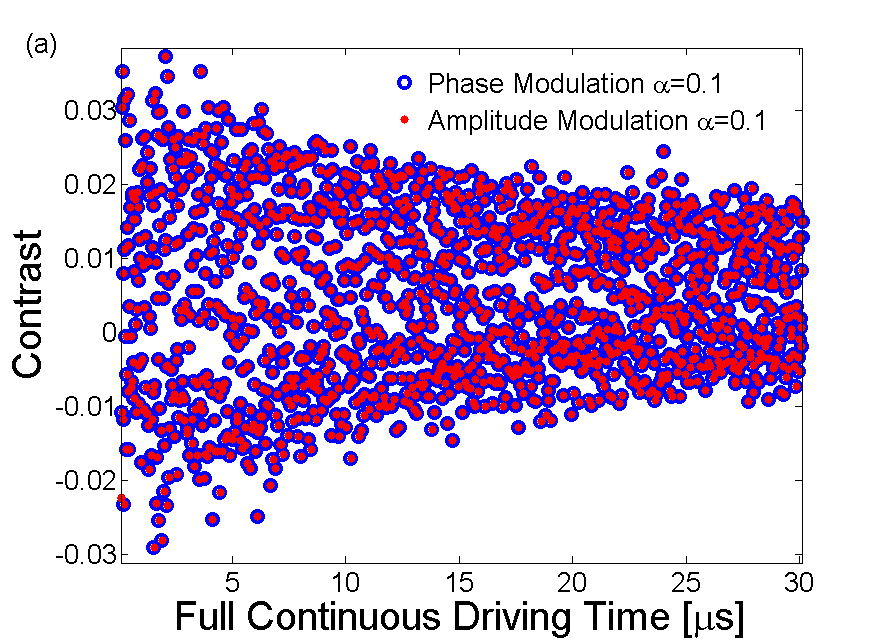} %fig3a.eps
	\includegraphics[width=1.05\columnwidth]{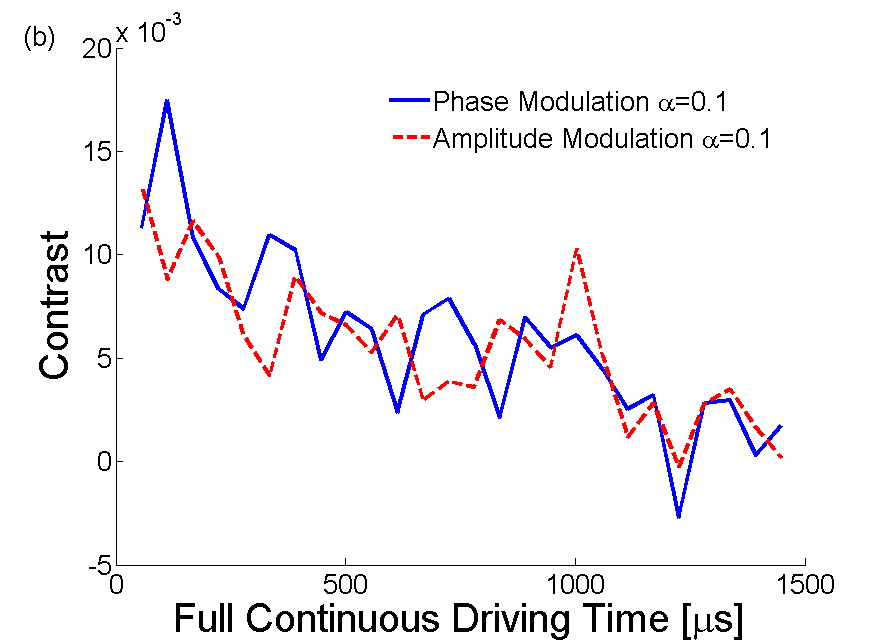} %fig3b.eps
	\caption{(Color online) Comparison between the fidelity of NV ensemble spin state initialized (a) along the $x$ axis and (b) along the $y$ axis for amplitude (Eq. \ref{equation:ampdetun}) and phase (Eq. \ref{equation:phasedetun}) modulations with equal strength $\alpha=0.1$. Other modulation strengths produced comparable results as well.} 
	\label{fig:ampvsphase}
\end{figure}

 In Fig. \ref{fig:ampvsphase} we demonstrate the similarity between the results obtained from amplitude (Eq. \ref{equation:ampdetun}) and phase (Eq. \ref{equation:phasedetun}) modulation with equal strength $\alpha=0.1$. Since both amplitude and phase were introduced by the same channels of the AWG acting as the same noise source, these results are comparable. Other modulation strengths provided comparable results as well. We expect that by using an AWG with faster sampling rate, allowing the full modulated waveform to be directly programmed, much higher phase accuracy could be achieved, thus providing an advantage to the phase-modulated scheme. 

\section{Appendix D: Description of fidelity measurements}
Our measurements presented in figure \ref{fig:driving} are performed as follows: we first use a long ($\sim 10 \mu$s) 532 nm laser pulse to initialize the ensemble state along the $z$ axis, measure the resulting fluorescence level $I_0$, and then rotate the state toward the $x$ axis using a resonant $\pi/2$ pulse. In order to extract the decoherence curves, we next need to measure the fidelity between this initial state and the final state after the modulated continuous driving is applied. In order to overcome technical drifts (such as instabilities from the laser / AOM), we project our final state twice using resonant $\pi/2$ pulses:  the first time toward the $z$ axis, and the second time toward the $-z$ axis,  and measure the resulting fluorescence levels $I_{f1},I_{f2}$ respectively. In order to express the fidelity between the final and initial states, these resulting fluorescence levels are then divided by the initial fluorescence level $r_1=I_{f1}/I_0,r_2=I_{f2}/I_0$, and in order to overcome the above-mentioned instabilities, the final results (the $y$ axis in the plots) is expressed in terms of contrast between these two values $C=\frac{r1-r2}{r1+r2}$.

\bibliography{nvbibliography}

\begin{thebibliography}{27}
\expandafter\ifx\csname natexlab\endcsname\relax\def\natexlab#1{#1}\fi
\expandafter\ifx\csname bibnamefont\endcsname\relax
  \def\bibnamefont#1{#1}\fi
\expandafter\ifx\csname bibfnamefont\endcsname\relax
  \def\bibfnamefont#1{#1}\fi
\expandafter\ifx\csname citenamefont\endcsname\relax
  \def\citenamefont#1{#1}\fi
\expandafter\ifx\csname url\endcsname\relax
  \def\url#1{\texttt{#1}}\fi
\expandafter\ifx\csname urlprefix\endcsname\relax\def\urlprefix{URL }\fi
\providecommand{\bibinfo}[2]{#2}
\providecommand{\eprint}[2][]{\url{#2}}

\bibitem[{\citenamefont{Taylor et~al.}(2008)\citenamefont{Taylor, Cappellaro,
  Childress, Jiang, Budker, Hemmer, Yacoby, Walsworth, and Lukin}}]{Taylor2008}
\bibinfo{author}{\bibfnamefont{J.~M.} \bibnamefont{Taylor}},
  \bibinfo{author}{\bibfnamefont{P.}~\bibnamefont{Cappellaro}},
  \bibinfo{author}{\bibfnamefont{L.}~\bibnamefont{Childress}},
  \bibinfo{author}{\bibfnamefont{L.}~\bibnamefont{Jiang}},
  \bibinfo{author}{\bibfnamefont{D.}~\bibnamefont{Budker}},
  \bibinfo{author}{\bibfnamefont{P.~R.} \bibnamefont{Hemmer}},
  \bibinfo{author}{\bibfnamefont{A.}~\bibnamefont{Yacoby}},
  \bibinfo{author}{\bibfnamefont{R.~L.} \bibnamefont{Walsworth}},
  \bibnamefont{and} \bibinfo{author}{\bibfnamefont{M.~D.} \bibnamefont{Lukin}},
  \bibinfo{journal}{Nat. Phys.} \textbf{\bibinfo{volume}{4}},
  \bibinfo{pages}{810} (\bibinfo{year}{2008}).

\bibitem[{\citenamefont{Maze et~al.}(2008)}]{Maze2008}
\bibinfo{author}{\bibfnamefont{J.~R.} \bibnamefont{Maze}} \bibnamefont{et~al.},
  \bibinfo{journal}{Nature (London)} \textbf{\bibinfo{volume}{455}},
  \bibinfo{pages}{644} (\bibinfo{year}{2008}).

\bibitem[{\citenamefont{Balasubramanian et~al.}(2008)}]{Balasubramanian2008}
\bibinfo{author}{\bibfnamefont{G.}~\bibnamefont{Balasubramanian}}
  \bibnamefont{et~al.}, \bibinfo{journal}{Nature (London)}
  \textbf{\bibinfo{volume}{455}}, \bibinfo{pages}{648} (\bibinfo{year}{2008}).

\bibitem[{\citenamefont{Grinolds et~al.}(2011)\citenamefont{Grinolds,
  Malentinsky, Hong, Lukin, Walsworth, and Yacoby}}]{Grinolds2011}
\bibinfo{author}{\bibfnamefont{M.~S.} \bibnamefont{Grinolds}},
  \bibinfo{author}{\bibfnamefont{P.}~\bibnamefont{Malentinsky}},
  \bibinfo{author}{\bibfnamefont{S.}~\bibnamefont{Hong}},
  \bibinfo{author}{\bibfnamefont{M.~D.} \bibnamefont{Lukin}},
  \bibinfo{author}{\bibfnamefont{R.~L.} \bibnamefont{Walsworth}},
  \bibnamefont{and} \bibinfo{author}{\bibfnamefont{A.}~\bibnamefont{Yacoby}},
  \bibinfo{journal}{Nat. Phys.} \textbf{\bibinfo{volume}{7}},
  \bibinfo{pages}{687} (\bibinfo{year}{2011}).

\bibitem[{\citenamefont{Pham et~al.}(2011)}]{Pham2011}
\bibinfo{author}{\bibfnamefont{L.~M.} \bibnamefont{Pham}} \bibnamefont{et~al.},
  \bibinfo{journal}{New J. Phys.} \textbf{\bibinfo{volume}{13}},
  \bibinfo{pages}{045021} (\bibinfo{year}{2011}).

\bibitem[{\citenamefont{Pham et~al.}(2012)\citenamefont{Pham, Bar-Gill,
  Belthangady, {Le Sage}, Cappellaro, Lukin, Yacoby, and Walsworth}}]{Pham2012}
\bibinfo{author}{\bibfnamefont{L.~M.} \bibnamefont{Pham}},
  \bibinfo{author}{\bibfnamefont{N.}~\bibnamefont{Bar-Gill}},
  \bibinfo{author}{\bibfnamefont{C.}~\bibnamefont{Belthangady}},
  \bibinfo{author}{\bibfnamefont{D.}~\bibnamefont{{Le Sage}}},
  \bibinfo{author}{\bibfnamefont{P.}~\bibnamefont{Cappellaro}},
  \bibinfo{author}{\bibfnamefont{M.~D.} \bibnamefont{Lukin}},
  \bibinfo{author}{\bibfnamefont{A.}~\bibnamefont{Yacoby}}, \bibnamefont{and}
  \bibinfo{author}{\bibfnamefont{R.~L.} \bibnamefont{Walsworth}},
  \bibinfo{journal}{Phys. Rev. B} \textbf{\bibinfo{volume}{86}},
  \bibinfo{pages}{045214} (\bibinfo{year}{2012}).

\bibitem[{\citenamefont{Acosta et~al.}(2009)}]{Acosta2009}
\bibinfo{author}{\bibfnamefont{V.~M.} \bibnamefont{Acosta}}
  \bibnamefont{et~al.}, \bibinfo{journal}{Phys. Rev. B}
  \textbf{\bibinfo{volume}{80}}, \bibinfo{pages}{115202}
  (\bibinfo{year}{2009}).

\bibitem[{\citenamefont{Acosta et~al.}(2010)\citenamefont{Acosta, Bauch,
  Jarmola, Zipp, Ledbetter, and Budker}}]{Acosta2010}
\bibinfo{author}{\bibfnamefont{V.~M.} \bibnamefont{Acosta}},
  \bibinfo{author}{\bibfnamefont{E.}~\bibnamefont{Bauch}},
  \bibinfo{author}{\bibfnamefont{A.}~\bibnamefont{Jarmola}},
  \bibinfo{author}{\bibfnamefont{L.~J.} \bibnamefont{Zipp}},
  \bibinfo{author}{\bibfnamefont{M.~P.} \bibnamefont{Ledbetter}},
  \bibnamefont{and} \bibinfo{author}{\bibfnamefont{D.}~\bibnamefont{Budker}},
  \bibinfo{journal}{Appl. Phys. Lett.} \textbf{\bibinfo{volume}{97}},
  \bibinfo{pages}{174104} (\bibinfo{year}{2010}).

\bibitem[{\citenamefont{Cappellaro and Lukin}(2009)}]{Cappellaro2009}
\bibinfo{author}{\bibfnamefont{P.}~\bibnamefont{Cappellaro}} \bibnamefont{and}
  \bibinfo{author}{\bibfnamefont{M.~D.} \bibnamefont{Lukin}},
  \bibinfo{journal}{Phys. Rev. A} \textbf{\bibinfo{volume}{80}},
  \bibinfo{pages}{032311} (\bibinfo{year}{2009}).

\bibitem[{\citenamefont{Bennett et~al.}(2013)\citenamefont{Bennett, Yao,
  Otterbach, Zoller, Rabl, and Lukin}}]{Bennett2013}
\bibinfo{author}{\bibfnamefont{S.~D.} \bibnamefont{Bennett}},
  \bibinfo{author}{\bibfnamefont{N.~Y.} \bibnamefont{Yao}},
  \bibinfo{author}{\bibfnamefont{J.}~\bibnamefont{Otterbach}},
  \bibinfo{author}{\bibfnamefont{P.}~\bibnamefont{Zoller}},
  \bibinfo{author}{\bibfnamefont{P.}~\bibnamefont{Rabl}}, \bibnamefont{and}
  \bibinfo{author}{\bibfnamefont{M.~D.} \bibnamefont{Lukin}},
  \bibinfo{journal}{Phys. Rev. Lett.} \textbf{\bibinfo{volume}{110}},
  \bibinfo{pages}{156402} (\bibinfo{year}{2013}).

\bibitem[{\citenamefont{Weimer et~al.}(2013)\citenamefont{Weimer, Yao, and
  Lukin}}]{Weimer2013}
\bibinfo{author}{\bibfnamefont{H.}~\bibnamefont{Weimer}},
  \bibinfo{author}{\bibfnamefont{N.~Y.} \bibnamefont{Yao}}, \bibnamefont{and}
  \bibinfo{author}{\bibfnamefont{M.~D.} \bibnamefont{Lukin}},
  \bibinfo{journal}{Phys. Rev. Lett.} \textbf{\bibinfo{volume}{110}},
  \bibinfo{pages}{067601} (\bibinfo{year}{2013}).

\bibitem[{\citenamefont{Ryan et~al.}(2010)\citenamefont{Ryan, Hodges, and
  Cory}}]{Ryan2010}
\bibinfo{author}{\bibfnamefont{C.~A.} \bibnamefont{Ryan}},
  \bibinfo{author}{\bibfnamefont{J.~S.} \bibnamefont{Hodges}},
  \bibnamefont{and} \bibinfo{author}{\bibfnamefont{D.~G.} \bibnamefont{Cory}},
  \bibinfo{journal}{Phys. Rev. Lett.} \textbf{\bibinfo{volume}{105}},
  \bibinfo{pages}{200402} (\bibinfo{year}{2010}).

\bibitem[{\citenamefont{Naydenov et~al.}(2011)\citenamefont{Naydenov, Dolde,
  Hall, Shin, Fedder, Hollenberg, Jelezko, and Wrachtrup}}]{Naydenov2011}
\bibinfo{author}{\bibfnamefont{B.}~\bibnamefont{Naydenov}},
  \bibinfo{author}{\bibfnamefont{F.}~\bibnamefont{Dolde}},
  \bibinfo{author}{\bibfnamefont{L.~T.} \bibnamefont{Hall}},
  \bibinfo{author}{\bibfnamefont{C.}~\bibnamefont{Shin}},
  \bibinfo{author}{\bibfnamefont{H.}~\bibnamefont{Fedder}},
  \bibinfo{author}{\bibfnamefont{L.~C.} \bibnamefont{Hollenberg}},
  \bibinfo{author}{\bibfnamefont{F.}~\bibnamefont{Jelezko}}, \bibnamefont{and}
  \bibinfo{author}{\bibfnamefont{J.}~\bibnamefont{Wrachtrup}},
  \bibinfo{journal}{Phys. Rev. B} \textbf{\bibinfo{volume}{83}},
  \bibinfo{pages}{081201(R)} (\bibinfo{year}{2011}).

\bibitem[{\citenamefont{Shim et~al.}(2012)\citenamefont{Shim, Niemeyer, Zhang,
  and Suter}}]{Shim2012}
\bibinfo{author}{\bibfnamefont{J.~H.} \bibnamefont{Shim}},
  \bibinfo{author}{\bibfnamefont{I.}~\bibnamefont{Niemeyer}},
  \bibinfo{author}{\bibfnamefont{J.}~\bibnamefont{Zhang}}, \bibnamefont{and}
  \bibinfo{author}{\bibfnamefont{D.}~\bibnamefont{Suter}},
  \bibinfo{journal}{Europhys. Lett.} \textbf{\bibinfo{volume}{99}},
  \bibinfo{pages}{40004} (\bibinfo{year}{2012}).

\bibitem[{\citenamefont{Hirose et~al.}(2012)\citenamefont{Hirose, Aiello, and
  Cappellaro}}]{Hirose2012}
\bibinfo{author}{\bibfnamefont{M.}~\bibnamefont{Hirose}},
  \bibinfo{author}{\bibfnamefont{C.~D.} \bibnamefont{Aiello}},
  \bibnamefont{and}
  \bibinfo{author}{\bibfnamefont{P.}~\bibnamefont{Cappellaro}},
  \bibinfo{journal}{Phys. Rev. A} \textbf{\bibinfo{volume}{86}},
  \bibinfo{pages}{062320} (\bibinfo{year}{2012}).

\bibitem[{\citenamefont{Cai et~al.}(2012)\citenamefont{Cai, Naydenov, Pfeiffer,
  McGuinness, Jahnke, Jelezko, Plenio, and Retzker}}]{Cai2012}
\bibinfo{author}{\bibfnamefont{J.}~\bibnamefont{Cai}},
  \bibinfo{author}{\bibfnamefont{B.}~\bibnamefont{Naydenov}},
  \bibinfo{author}{\bibfnamefont{R.}~\bibnamefont{Pfeiffer}},
  \bibinfo{author}{\bibfnamefont{L.}~\bibnamefont{McGuinness}},
  \bibinfo{author}{\bibfnamefont{K.}~\bibnamefont{Jahnke}},
  \bibinfo{author}{\bibfnamefont{F.}~\bibnamefont{Jelezko}},
  \bibinfo{author}{\bibfnamefont{M.}~\bibnamefont{Plenio}}, \bibnamefont{and}
  \bibinfo{author}{\bibfnamefont{A.}~\bibnamefont{Retzker}},
  \bibinfo{journal}{New J. Phys.} \textbf{\bibinfo{volume}{14}},
  \bibinfo{pages}{113023} (\bibinfo{year}{2012}).

\bibitem[{\citenamefont{Farfurnik et~al.}(2015)\citenamefont{Farfurnik,
  Jarmola, Pham, Wang, Dobrovitski, Walsworth, Budker, and
  Bar-Gill}}]{Farfurnik2015}
\bibinfo{author}{\bibfnamefont{D.}~\bibnamefont{Farfurnik}},
  \bibinfo{author}{\bibfnamefont{A.}~\bibnamefont{Jarmola}},
  \bibinfo{author}{\bibfnamefont{L.}~\bibnamefont{Pham}},
  \bibinfo{author}{\bibfnamefont{Z.}~\bibnamefont{Wang}},
  \bibinfo{author}{\bibfnamefont{V.}~\bibnamefont{Dobrovitski}},
  \bibinfo{author}{\bibfnamefont{R.}~\bibnamefont{Walsworth}},
  \bibinfo{author}{\bibfnamefont{D.}~\bibnamefont{Budker}}, \bibnamefont{and}
  \bibinfo{author}{\bibfnamefont{N.}~\bibnamefont{Bar-Gill}},
  \bibinfo{journal}{Phys. Rev. B} \textbf{\bibinfo{volume}{92}},
  \bibinfo{pages}{060301(R)} (\bibinfo{year}{2015}).

\bibitem[{\citenamefont{Teissier et~al.}(2017)\citenamefont{Teissier, Barfuss,
  and Maletinsky}}]{Teissier2017}
\bibinfo{author}{\bibfnamefont{J.}~\bibnamefont{Teissier}},
  \bibinfo{author}{\bibfnamefont{A.}~\bibnamefont{Barfuss}}, \bibnamefont{and}
  \bibinfo{author}{\bibfnamefont{P.}~\bibnamefont{Maletinsky}},
  \bibinfo{journal}{Journal of Optics} \textbf{\bibinfo{volume}{19}},
  \bibinfo{pages}{044003} (\bibinfo{year}{2017}).

\bibitem[{\citenamefont{Solomon}(1959)}]{Solomon1959}
\bibinfo{author}{\bibfnamefont{I.}~\bibnamefont{Solomon}},
  \bibinfo{journal}{Phys. Rev. Lett.} \textbf{\bibinfo{volume}{2}},
  \bibinfo{pages}{301} (\bibinfo{year}{1959}).

\bibitem[{\citenamefont{Timoney et~al.}(2011)\citenamefont{Timoney, Baumgart,
  Johanning, Varón, Plenio, Retzker, and Wunderlich}}]{Timoney2011}
\bibinfo{author}{\bibfnamefont{N.}~\bibnamefont{Timoney}},
  \bibinfo{author}{\bibfnamefont{I.}~\bibnamefont{Baumgart}},
  \bibinfo{author}{\bibfnamefont{M.}~\bibnamefont{Johanning}},
  \bibinfo{author}{\bibfnamefont{A.~F.} \bibnamefont{Varón}},
  \bibinfo{author}{\bibfnamefont{M.~B.~.} \bibnamefont{Plenio}},
  \bibinfo{author}{\bibfnamefont{A.}~\bibnamefont{Retzker}}, \bibnamefont{and}
  \bibinfo{author}{\bibfnamefont{C.}~\bibnamefont{Wunderlich}},
  \bibinfo{journal}{Nature (London)} \textbf{\bibinfo{volume}{476}},
  \bibinfo{pages}{185} (\bibinfo{year}{2011}).

\bibitem[{\citenamefont{Aharon et~al.}(2013)\citenamefont{Aharon, Drewsen, and
  Retzker}}]{Aharon2013}
\bibinfo{author}{\bibfnamefont{N.}~\bibnamefont{Aharon}},
  \bibinfo{author}{\bibfnamefont{M.}~\bibnamefont{Drewsen}}, \bibnamefont{and}
  \bibinfo{author}{\bibfnamefont{A.}~\bibnamefont{Retzker}},
  \bibinfo{journal}{Phys. Rev. Lett.} \textbf{\bibinfo{volume}{111}},
  \bibinfo{pages}{230507} (\bibinfo{year}{2013}).

\bibitem[{\citenamefont{Cohen et~al.}(2016)\citenamefont{Cohen, Aharon, and
  Retzker}}]{Cohen2016}
\bibinfo{author}{\bibfnamefont{I.}~\bibnamefont{Cohen}},
  \bibinfo{author}{\bibfnamefont{N.}~\bibnamefont{Aharon}}, \bibnamefont{and}
  \bibinfo{author}{\bibfnamefont{A.}~\bibnamefont{Retzker}},
  \bibinfo{journal}{Fortschr. Phys.} \textbf{\bibinfo{volume}{64}},
  \bibinfo{pages}{1} (\bibinfo{year}{2016}).

\bibitem[{\citenamefont{Aharon et~al.}(2016)\citenamefont{Aharon, Cohen,
  Jelezko, and Retzker}}]{Aharon2016}
\bibinfo{author}{\bibfnamefont{N.}~\bibnamefont{Aharon}},
  \bibinfo{author}{\bibfnamefont{I.}~\bibnamefont{Cohen}},
  \bibinfo{author}{\bibfnamefont{F.}~\bibnamefont{Jelezko}}, \bibnamefont{and}
  \bibinfo{author}{\bibfnamefont{A.}~\bibnamefont{Retzker}},
  \bibinfo{journal}{New J. Phys.} \textbf{\bibinfo{volume}{18}},
  \bibinfo{pages}{123012} (\bibinfo{year}{2016}).

\bibitem[{\citenamefont{Aiello et~al.}(2013)\citenamefont{Aiello, Hirose, and
  Cappellaro}}]{Aiello2013}
\bibinfo{author}{\bibfnamefont{C.~D.} \bibnamefont{Aiello}},
  \bibinfo{author}{\bibfnamefont{M.}~\bibnamefont{Hirose}}, \bibnamefont{and}
  \bibinfo{author}{\bibfnamefont{P.}~\bibnamefont{Cappellaro}},
  \bibinfo{journal}{Nat. Commun.} \textbf{\bibinfo{volume}{4}},
  \bibinfo{pages}{1419} (\bibinfo{year}{2013}).

\bibitem[{\citenamefont{Wang and Uhlenbeck}(1945)}]{Wang1945}
\bibinfo{author}{\bibfnamefont{M.~C.} \bibnamefont{Wang}} \bibnamefont{and}
  \bibinfo{author}{\bibfnamefont{G.~E.} \bibnamefont{Uhlenbeck}},
  \bibinfo{journal}{Rev. Mod. Phys.} \textbf{\bibinfo{volume}{17}},
  \bibinfo{pages}{323} (\bibinfo{year}{1945}).

\bibitem[{\citenamefont{Hanson et~al.}(2008)\citenamefont{Hanson, Dobrovitski,
  Feiguin, and Gywat}}]{Hanson2008}
\bibinfo{author}{\bibfnamefont{R.}~\bibnamefont{Hanson}},
  \bibinfo{author}{\bibfnamefont{V.~V.} \bibnamefont{Dobrovitski}},
  \bibinfo{author}{\bibfnamefont{A.~E.} \bibnamefont{Feiguin}},
  \bibnamefont{and} \bibinfo{author}{\bibfnamefont{O.}~\bibnamefont{Gywat}},
  \bibinfo{journal}{Science} \textbf{\bibinfo{volume}{320}},
  \bibinfo{pages}{352} (\bibinfo{year}{2008}).

\bibitem[{\citenamefont{Gillespie}(1996)}]{Gillespie1996}
\bibinfo{author}{\bibfnamefont{D.~T.} \bibnamefont{Gillespie}},
  \bibinfo{journal}{Phys. Rev. E.} \textbf{\bibinfo{volume}{54}},
  \bibinfo{pages}{2084} (\bibinfo{year}{1996}).

\end{thebibliography}

\end{document}